\documentclass[%
 aip,
 amsmath,amssymb,
 preprint,%
]{revtex4-1}

\usepackage{graphicx}
\usepackage{dcolumn}
\usepackage{bm}

\usepackage[utf8]{inputenc}
\usepackage[T1]{fontenc}
\usepackage{mathptmx}
\usepackage{etoolbox}
\usepackage{subcaption}
\usepackage{color}
\usepackage{soul}

\makeatletter
\def\@email#1#2{%
 \endgroup
 \patchcmd{\titleblock@produce}
  {\frontmatter@RRAPformat}
  {\frontmatter@RRAPformat{\produce@RRAP{*#1\href{mailto:#2}{#2}}}\frontmatter@RRAPformat}
  {}{}
}%
\makeatother

\renewcommand\hl[1]{#1} 

\begin{document}

\title{EUV Debris Mitigation using Magnetic Nulls}

\author{B. Y. Israeli}
\affiliation{Princeton Plasma Physics Laboratory, Princeton, NJ, USA}
\affiliation{Princeton University, Princeton, NJ, USA}
\email{bisraeli@pppl.gov}
\author{C. B. Smiet}
\affiliation{Princeton Plasma Physics Laboratory, Princeton, NJ, USA}
\author{M. Simeni Simeni}
\affiliation{Department of Mechanical Engineering, University of Minnesota, Minneapolis, MN, USA}
\affiliation{Princeton Plasma Physics Laboratory, Princeton, NJ, USA}
\author{A. Diallo}
\affiliation{Princeton Plasma Physics Laboratory, Princeton, NJ, USA}

\date{\today}
\begin{abstract}
Next generation EUV sources for photolithography use light produced by laser-produced plasmas (LPP) from ablated tin droplets. 
A major challenge for \hl{extending the lifetime of} these devices is mitigating damage caused by deposition of tin debris on the sensitive collection mirror. 
Especially difficult to stop are high energy (up to 10 keV) highly charged tin ions created in the plasma. 
Existing solutions include the use of stopping gas, electric fields, and magnetic fields. 
One common configuration consists of a magnetic field perpendicular to the EUV emission direction, but such a system can \hl{result in ion populations that are trapped rather than removed}.
We investigate a \hl{previously unconsidered} mitigation geometry consisting of a magnetic null by performing full-orbit integration of the ion trajectories in an EUV system with realistic dimensions, and optimize the coil locations for the null configuration. 
The magnetic null prevents a fraction of ions from hitting the mirror comparable to that of the perpendicular field, but does not trap any ions due to \hl{the chaotic nature of ion trajectories that} pass close to the null.  
This technology can potentially improve LPP-based EUV photolithography system efficiency and lifetime, and \hl{may allow} for a different, more efficient formulation of buffer gas.
\end{abstract}

\maketitle

Debris mitigation is one of the main challenges for furthering the development of reliable molten tin droplet-based EUV lithography sources. Upon drive laser-tin droplet interaction, slow and fast tin ionic products as well as neutral atoms (within molten tin splashes) are generated. These constitute the so-called “debris”, which are responsible for damage to the delicate Mo/Si multilayer Bragg reflectors used for the collection of the produced EUV emission (at 13.5$\pm$ 1$\%$ nm). Despite the fact that ASML (for instance) managed to reduce collector mirror degradation rate (due to debris) down to 0.05\% per giga-pulse \cite{verhoeven20200}, \hl{there remains substantial room for innovation} to extend the lifetime of EUV systems \cite{tomie2012tin, takahashi2008emission}.

Recently, effective mitigation strategies for molten tin splashes and low energy\hl{ }tin ions have been demonstrated \cite{nakamura2007mitigation, klunder2005debris, harilal2007ion, elg2016situ, abramenko2018measurements, wu2012debris, sporre2016modeling, sizyuk2017background, stodolna2018controlling}. These mitigation strategies mainly consist in the introduction of up to a few millibar of H$_{2}$ background gas. H$_2$ serves two specific purposes: (1) It slows down low energy ions, reducing (through collisions) their kinetic energy before their impact on the mirror surfaces and (2) It cleans up the tin splashes (deposits) on the mirror surfaces through the chemical reaction: Sn (s) + 4H (g) → SnH$_4$ (g). The end molecule SnH$_4$ (stannane), being gaseous, can be detached from the surfaces and pumped out of the vessel using widely-available vacuum systems. \hl{The buffer gas also has the effect of reducing the charge state of Sn ions via electron capture from neutral H$_2$.}\cite{raiEvidenceProductionKeV2023,abramenko2018measurements} 

The use of a background gas also has significant downsides: Energetic H ions can implant in the mirror and aggregate between the Mo/Si layers causing blistering and delamination~\cite{kuznetsov2013ion, tomuro2022evaluation}. 
Additionally, the aforementioned (stopping background gas shield) strategy cannot be extrapolated \hl{indefinitely} to the mitigation of fast\hl{ }ionic products since a significantly higher background gas pressure would be required for a substantial ion kinetic energy reduction. \hl{A} significant increase of the H$_2$ background pressure has detrimental effects on the EUV collection efficiency following EUV absorption by the background gas \cite{banine2011physical, harilal2007ion, wu2012debris}.
These detrimental effects are all reduced by a lower density background gas, and the composition of this gas needs to be optimized for the specific device and with respect to other mitigation methods used.

A method that can complement the gas shield and prevent high-energy Sn ions from hitting the mirror is magnetic debris mitigation.
Magnetic debris mitigation is based on the principle that charged particles are guided along magnetic field lines in a helical trajectory. 
In this way, the harmful Sn ions can be guided away from the mirror, and onto parts of the device that act as cold traps, removing them from the system. A simple magnetic debris mitigation geometry\hl{, consisting} of two coils outside of the EUV system, creat\hl{es} a magnetic field perpendicular to the direction of light emission. 
This geometry is actively investigated~\cite{harilal2007ion, ueno2008magnetic, elg2015magnetic, hosoda2021development}, and is patented by Gigaphoton Inc\hl{.}~\cite{hoshino2006light} 
We call this configuration the \emph{perpendicular field}.

\hl{While the helical paths of charged particles generally align to magnetic field lines, a non-constant field strength along these trajectories can }
\hl{Provided the magnetic field never vanishes and the field strength changes slowly along the trajectory of a charged particle, the magnetic moment of the particle is conserved, a property known as 'adiabatic invariance'.}
A charged particle moving along a helical path parallel to a magnetic field into a region of stronger magnetic field\hl{ }must \hl{therefore} increase its\hl{ }velocity \hl{perpendicular to the field }$v_\perp$\hl{.} 
The parallel velocity $v_{||}$ decreases \hl{by energy conservation}, and in a sufficiently strong field the particle is reflected back. \hl{Particles can therefore be trapped on trajectories that oscillate between local maxima of the magnetic field strength.}
This mechanism is used for confinement in nuclear fusion devices called magnetic mirrors \cite{post1987magnetic}. 

Though the goal of debris mitigation and magnetic confinement are opposite, the governing equations are identical. 
Magnetic mirrors will always lose a fraction of the particles confined, which can be expressed in terms of the mirror ratio $r_{\rm mirror}={B_{\rm max}}/{B_{\rm min}}$.
Particles where ${v_{||}}/{v_\perp} > {\sqrt{r_{\rm mirror}-1}}$
are are in the 'loss cone' and will escape the mirror; all other particles are trapped in the mirror. 

\hl{These trapped ions present a notable problem for debris mitigation using magnetic fields. Optimally, a magnetic mitigation scheme would direct ions away from the mirror and towards designated target surfaces within a short travel distance. Longer ion trajectories may be expected to increase the likelihood of charge exchange with the buffer gas and scattering against both the buffer gas and other ions. Charge exchange (and eventual neutralization) reduce (and eventually remove) the ability of the magnetic field to confine and direct the ions, resulting in uncontrolled deposition of debris. A magnetic field that traps ions therefore cannot easily mitigate the impact of these ions on the mirror. The details of this effect and of ion diffusion due to scattering are inherently dependent on buffer gas parameters and are left outside the scope of this preliminary work.}


In an EUV debris mitigation system the ratio of ${B_{\rm max}}/{B_{\rm min}}$ can be varied. \hl{For a fixed distance between coils, this ratio is inversely dependent on coil radius.}
A Helmholtz configuration\hl{ (coil diameter equal to coil separation), in which the central minimum of $|B|$ is removed, }would have minimal trapping. \hl{However, this }requires very large coils, stored magnetic energy, and therefore cost of the system,\hl{ suggesting the use of smaller coils which produce some trapping.}

\hl{I}n the mirror configuration, ions are trapped because of adiabatic invariance, but this invariance is lost when the particle reaches a point of very low magnetic field.
In this paper we will therefore consider a magnetic configuration that does not exhibit trapping for the purposes of debris mitigation, specifically a configuration containing a magnetic null~\cite{parnell1996structure}.

\begin{figure}
    \includegraphics[width=0.6\linewidth]{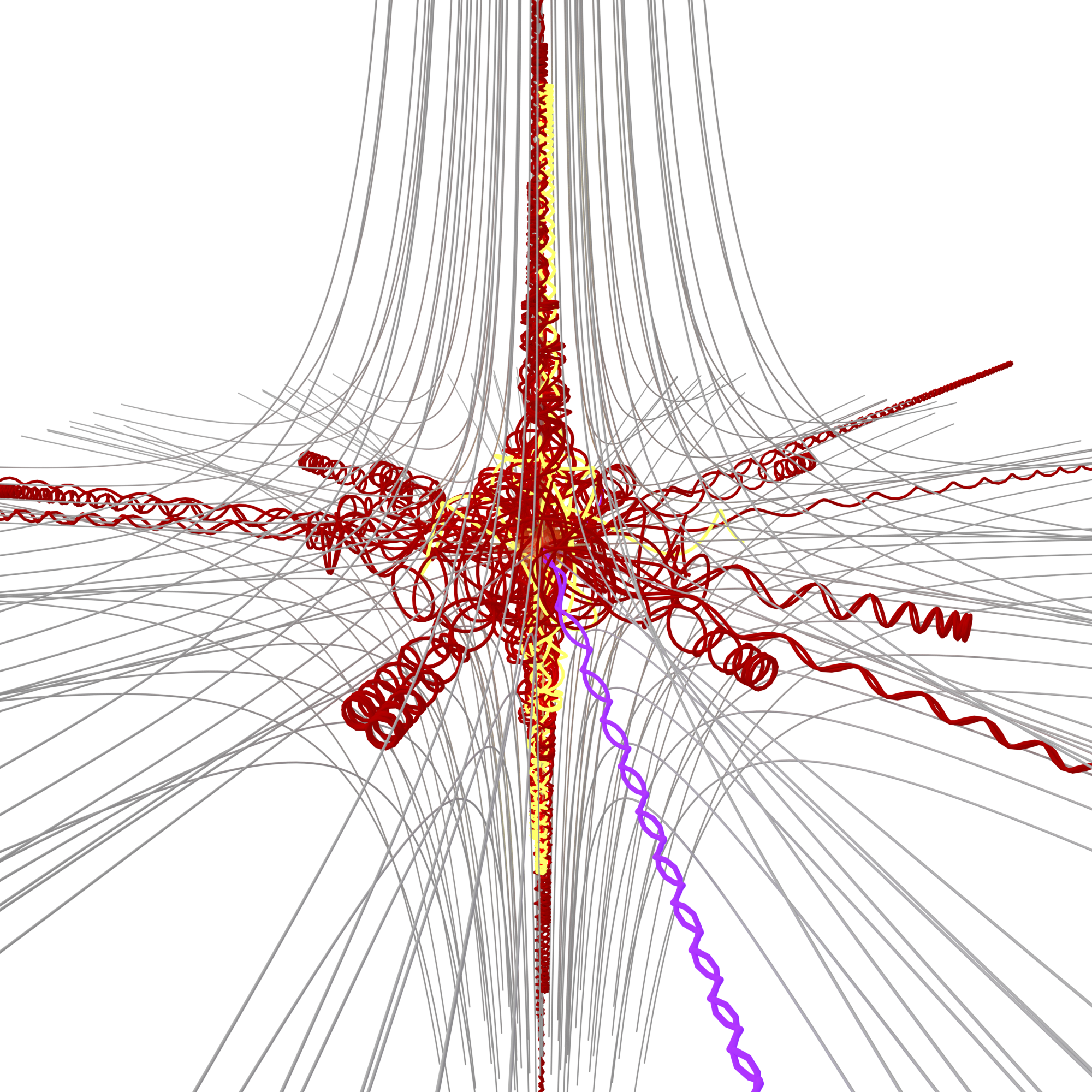}
    \caption{\label{fig:null_trajectories} Ion trajectories around \hl{a simple} magnetic null. \hl{The magnetic field is given by $\mathbf{B} = B_0(\hat{x} +
    \hat{y} - 2\hat{z})$ and magnetic field lines are shown in gray.} The fan plane is horizontal, and the spine is along the \hl{z}-axis. \hl{15} ion trajectories starting at the null are shown. \hl{Of these, 13 are colored red, one is yellow and one is purple. The ions travel out over the fan plane or along the spine until they bounce back, and lie on erratic trajectories. Particles whose velocity is closely aligned with the magnetic field (purple trajectory) travel much further before bouncing back.}}
\end{figure}

Magnetic nulls, or cusps, have been considered for confinement of fusion plasmas as well, for example in the bionic cusp developed by Grad \cite{grad1961containment} and later designs such as the polywell~\cite{keller1966confinement}.
This approach is mostly abandoned because it does not exhibit good confinement. 
A magnetic null consists of a fan plane and a spine as illustrated in figure~\ref{fig:null_trajectories}. 
Ion trajectories around a null are erratic~\cite{grad1961containment}, as each time they pass by the null they pick up a new random pitch angle ($v_{||}/v_\perp$)\hl{ and a different magnetic moment}.

A null configuration can be created with just two coils creating magnetic fields in opposing directions, which cancel each other in the center to create the null. 
If the null is placed where the tin ions are generated, their trajectories will be erratic, and they will eventually leave the system. 
In this paper we will investigate this null magnetic configuration for the purposes of debris mitigation in EUV systems by integrating the trajectories of a population of EUV ions. 
We will compare it to the perpendicular field system with similar parameters that are achievable with commercially available High Temperature Superconductor (HTS) tapes up to 2T.

We will investigate two families of magnetic configurations, the\hl{ }magnetic null configuration, and a perpendicular field configuration akin to the configurations described in \cite{ueno2008magnetic, elg2015magnetic, harilal2007ion, hosoda2021development}.
In the magnetic null configuration we will vary the vertical position of the coils as well as the field strength, in the perpendicular field we will vary the field strength. 

\begin{figure}
    \centering
    \includegraphics[width=\linewidth]{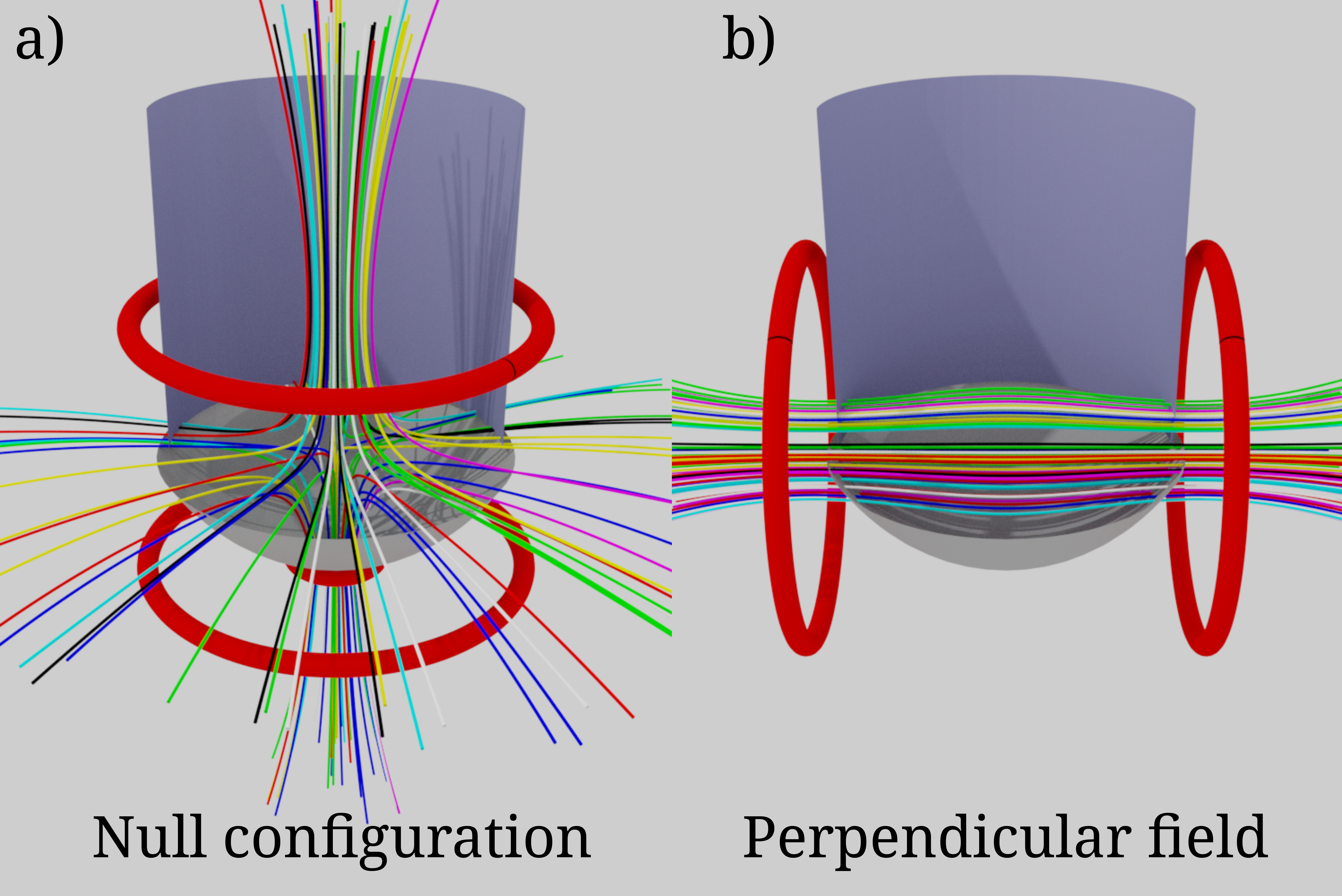}
    \caption{\label{fig:device geometry}Device geometry of the two magnetic debris mitigation configurations considered. Coils are shown in red, the elliptic mirror in grey, and a cylindrical side wall in blue. Not shown is a top plate. Select magnetic field lines are plotted\hl{ in various colors} to demonstrate the magnetic configuration in a: the null configuration and b: the perpendicular field configuration.}
\end{figure}

The model EUV system geometry is based on the publicly available parameters of the EUV system utilized by ASML in their next-generation photolithography machines. 
The simulated geometry consists of a cylindrical wall, capped with an elliptical mirror at one end and a flat plate at the other end. A cutaway view which includes magnetic field lines is shown in Figure \ref{fig:device geometry}. The cylinder is $30$ cm in radius and has its axis passing through the origin. The mirror is elliptical with one focus at the origin, the center of the mirror located 20 cm from this focus, an inter-focal distance of 60 cm, and the axis coinciding with that of the cylinder. The mirror has a 6 cm diameter circular hole at its center for the excitation laser to pass through. The plate is located opposite to the mirror, 60 cm from the origin, normal to the device axis. With this choice of parameters, the mirror intersects the cylindrical wall at z=-3.07 cm. We will work in cylindrical coordinates, with the z-axis coinciding with the device axis and the mirror located at negative z.
Ions originate at the origin, the focus of the elliptic mirror.

The magnetic null field configuration \hl{is generated with three coils: two large primary coils create the magnetic null geometry, and a smaller tertiary coil (coil 3) is placed just below the mirror.} 
The two primary coils have a radius of 35 cm. Their axial locations and currents are taken to be variable. We call the coil closer to the mirror coil 1 and the coil further from the mirror coil 2. \hl{The} tertiary coil of 8 cm radius is placed below the mirror, at z=-22 cm. Its field is fixed at 0.5 T, pointed toward positive z. 
\hl{The magnetic field of the coils is calculated using the Biot-Savart Law, and the coils are approximated as filamentary, consisting of 1000 piecewise linear segments.}  
\hl{The field created by these coils is shown in} Figure \ref{fig:device geometry}\hl{ a}. Coils 1 and 2  create the majority of the field, and coil 3 pinches the magnetic field lines closer together at the mirror. This additional coil, \hl{though not necessary for the creation of the null geometry, was found to enhance the mitigation by concentrating the strike points of the ions into the hole in the mirror}.

The variable parameters defining the magnetic field geometry are the axial locations and currents/fields of the two primary coils. 
We vary the axial locations of both coils and the field of coil 1. The field of coil 2 (B$_2$) will be constrained to place a magnetic null at the origin. The location of coil 1 is varied from z$_1$=-22 cm to z$_1$= 0 cm in 5 steps, and the position of coil 2 is varied between z$_2$ =0 cm and z$_2$ =50 cm in 5 steps. The field of coil 1 is varied between B$_1$= 0T and B$_1$=2T (pointing toward positive z) in 10 steps. If the field magnitude of coil 2 required to produce a null at the origin exceeds 2T, the configuration is discarded. Of the 250 possible location and field combinations, 113 had an acceptable coil 2 field strength and were used.

The perpendicular field configuration consists of two coils, creating a magnetic field perpendicular to the $z$-axis. The same wall and mirror geometry and the same ion populations as in the magnetic null simulations are used. The coils in this test configuration have the same radius as those in the null configuration, of 35 cm and are 35 cm from the origin on either side of the device. Field lines of this configuration are shown in Figure \ref{fig:device geometry} b. The magnetic field strength is lower in the center, as is seen from the bulging of the field lines. This can cause trapping of ions, which can be reflected from the higher field regions near the coils.

Discrete tin ion populations with fixed energy and charge state were considered. 
The populations had energies between 1 eV and 10 keV, divided logarithmically in 10 steps, and with all integer charge states between +1 and +8. For each coil configuration, the trajectories of 256 ions from each population were integrated, totalling 20,480 trajectories per configuration.

\begin{figure}
    \centering
    \includegraphics[width=\linewidth]{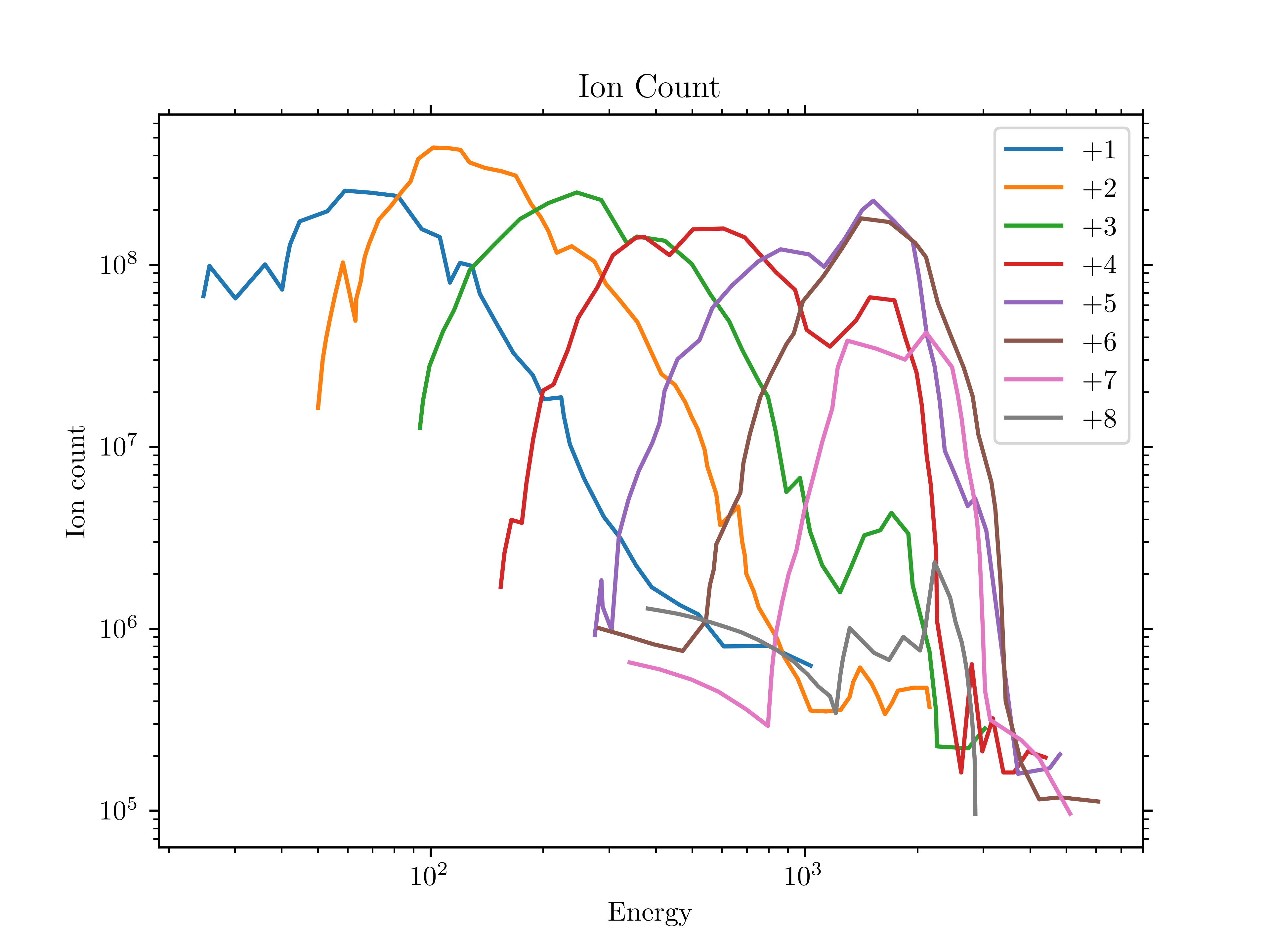}
    \caption{\label{fig:ion dist}The ion distribution with respect to energy in eV (horizontal axis) and charge state (in legend) reproduced from Versolato et al.~\cite{versolato2019physics}.}
\end{figure}

In order to consider a distribution of ion energies and charges that might\hl{ }reflect the ion flux in a real device, ion counts were normalized with respect to a distribution from~\cite{versolato2019physics} during analysis. This distribution is shown in Figure \ref{fig:ion dist}. Since these measurements do not encompass the whole energy \hl{range} considered at all charge states, the ion count was assumed to be zero outside the measured ranges.\hl{ Experiments have shown that charge exchange with buffer gas can reduce the charge states of Sn ions, producing a flux of largely Sn$^+$ and Sn$^{2+}$. This occurs without an appreciable reduction in ion energy.}\cite{raiEvidenceProductionKeV2023} \hl{As such, calculations were also performed assuming all ions are replaced with Sn$^+$. This was accomplished by summing the distributions in figure} \ref{fig:ion dist}\hl{, normalizing the distribution of Sn$^+$ with respect to this total, and discarding all other charge states. This makes the non-physical assumption that all ions undergo electron capture to Sn$^+$ immediately, but functions as a test of the maximal possible impact of this effect. A more detailed analysis with electron capture occurring stochastically over the course of ion trajectories is left for later work.}

The ion trajectories were integrated by numerically solving the Lorentz equation as two coupled first-order differential equations for the particle velocity $\mathbf{v}$ and position $\mathbf{x}$:
\begin{align}
  \frac{\partial \mathbf{v}}{\partial t} &= \frac{q_{\rm ion}}{m_{\rm ion}} (\mathbf{v} \times \mathbf{B}(\mathbf{x})) \\
  \frac{\partial \mathbf{x}}{\partial t} &= \mathbf{v}
\end{align}
where $q_{\rm ion}$ is the charge of the ion,  $m_{\rm ion}$ is the mass of a tin ion and $\mathbf{B}$ is the magnetic field vector, calculated using the Biot-Savart Law.\hl{ This was considered a reasonable approximation of ion trajectories over the distances considered (as opposed to the hydrodynamic models applied to earlier stages of the plasma expansion}\cite{hemmingaHighenergyIonsNd2021}\hl{), as the ion-ion and ion-electron mean free paths can be expected to be much larger than the plasma diameter once the plasma has reached several centimeters in size}\cite{brandstatterTemporallySpatiallyResolved2018}.
The electric term in the Lorentz equation is ignored, as the electric field is assumed negligible on the whole device scale. 
The integration is performed using the DOP853 integration routine implementing an 8th order Runge-Kutta method, and the equations are solved in Python using just-in-time compilation with Numba for evaluation of the Biot-Savart integral. 
The initial condition for the position $\mathbf{x}$ is the origin, and the velocity magnitude $|v|$ is specified by the ion energy through $|v| = \sqrt{2E/m_{ion}}$. 
The direction of the velocity is randomly chosen from a uniform spherical distribution. Each trajectory is integrated until either it intersects the wall or mirror, or it reaches a fixed length, set to $5$ meters in the case of the parameter scan.

\begin{figure}
    \centering
    \includegraphics[width=\linewidth]{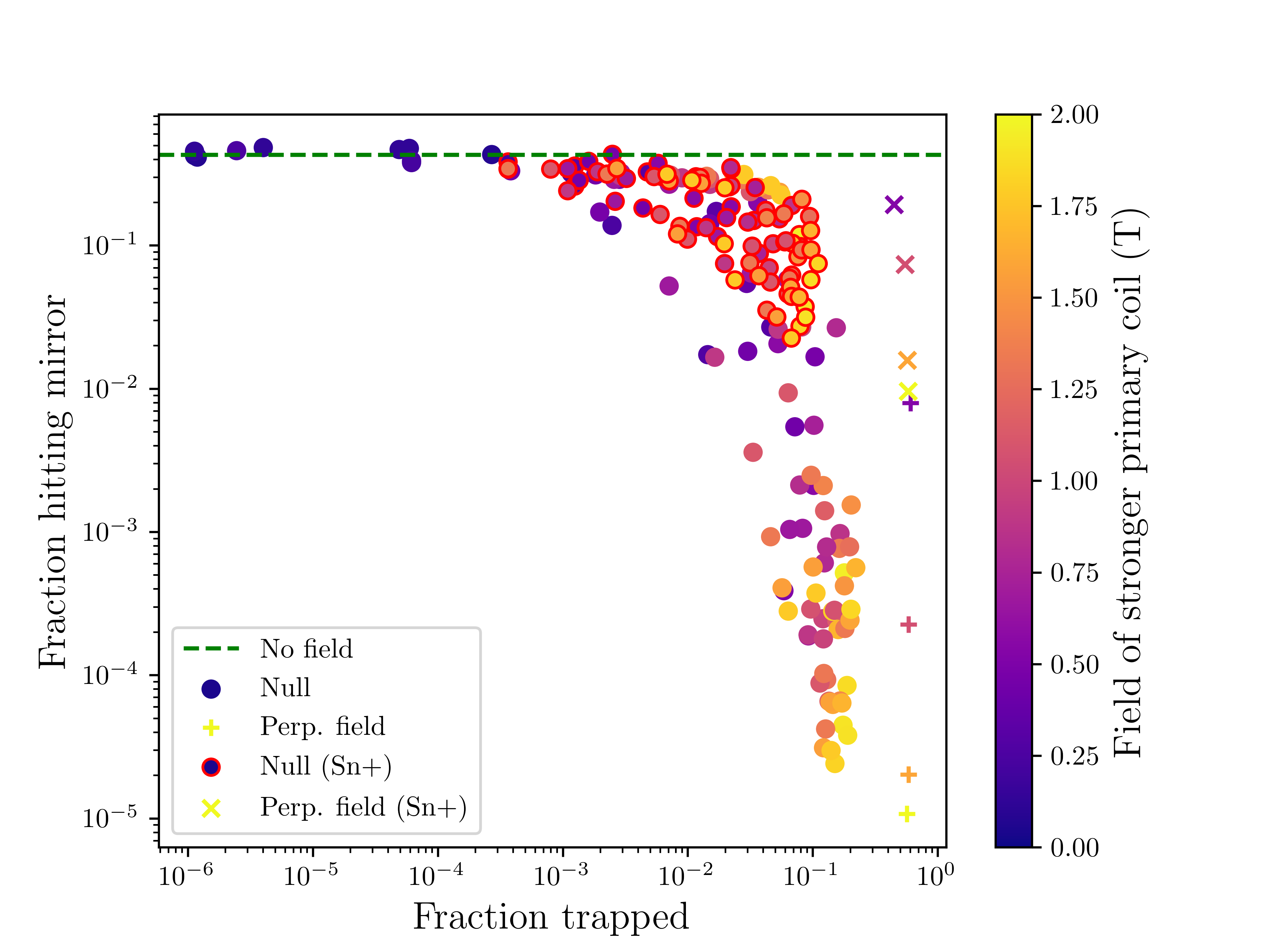}
    \caption{Plot of the \hl{fraction} of trapped ions and \hl{fraction} of ions which hit the mirror for each null configuration tested in the parameter scan. \hl{P}erpendicular field configurations with varied field strength (between 0.5 T and 2.0 T) are shown for comparison.\hl{ Also shown are fractions assuming all ions are replaced with Sn$^+$ as a test of the effect of charge exchange with H$_2$. }The field strength produced by the stronger of the two primary coils (or that of either coil in the perpendicular field case) is indicated by color. The fraction of ions hitting the mirror with no magnetic field is shown with a green line.\label{fig:scatter}}
    
\end{figure}

The results of the parameter scan are shown in figure \ref{fig:scatter}. An inverse relation between trapping and mitigation is visible, with a reduction in ion flux to the mirror corresponding to an apparent increase in the number of ions trapped in the null for longer than the integration length. As might be expected, stronger fields are seen to produce greater mitigation, but also greater trapping (due to mirroring effects). It can be seen that the perpendicular field can produce a comparable or smaller ion flux to the mirror, again correlating with field strength, but with substantially higher trapping. As discussed above, the trapping rate of the perpendicular field is determined by coil geometry and is independent of field strength. \hl{The case with all ions replaced with Sn$^+$ follows a similar pattern, but with increased fractions of the ions reaching the mirror. It can be inferred that, with sufficiently high buffer gas pressure, charge exchange may be expected to reduce but not eliminate the efficacy of magnetic debris mitigation techniques.}

In order to further investigate the comparative performance of these two magnetic field topologies, we selected the null configuration from the parameter scan which had the lowest ion flux to the mirror for further study, comparing it against the perpendicular field configuration with a similar (2.0 T) field strength. The null configuration had coil 1 located at z=-22 cm and coil 2 at z=25 cm. These configurations, and their performance are detailed in table \ref{tab:results}.
\begin{table}[h]
    \centering
    \begin{tabular}{r|c|c|c}
        &  Null & Perp. field & No field\\
       \hline
       Coil fields (T) & 1.56 (coil 1) & 2.0 & N/A \\
        & 1.83 (coil 2) & 2.0 & \\
        & 0.50 (coil 3) & & \\
       Ion fraction hitting mirror (\%) & 0.0039 & 0.00040 & 42 \\
       Ion fraction trapped (\%) & 2.1 & 58 & 0.0 
    \end{tabular}
    \caption{Parameters for the compared null and perpendicular field configurations. A configuration with no magnetic field is included as a baseline. Ion fractions are indicated for an integration length of 10 m. The trapped fraction for the perpendicular field matches the analytically calculated trapped fraction for a mirror of this geometry (59\%).}
    \label{tab:results}
\end{table}

\begin{figure}
    \centering
    \includegraphics[width=\linewidth]{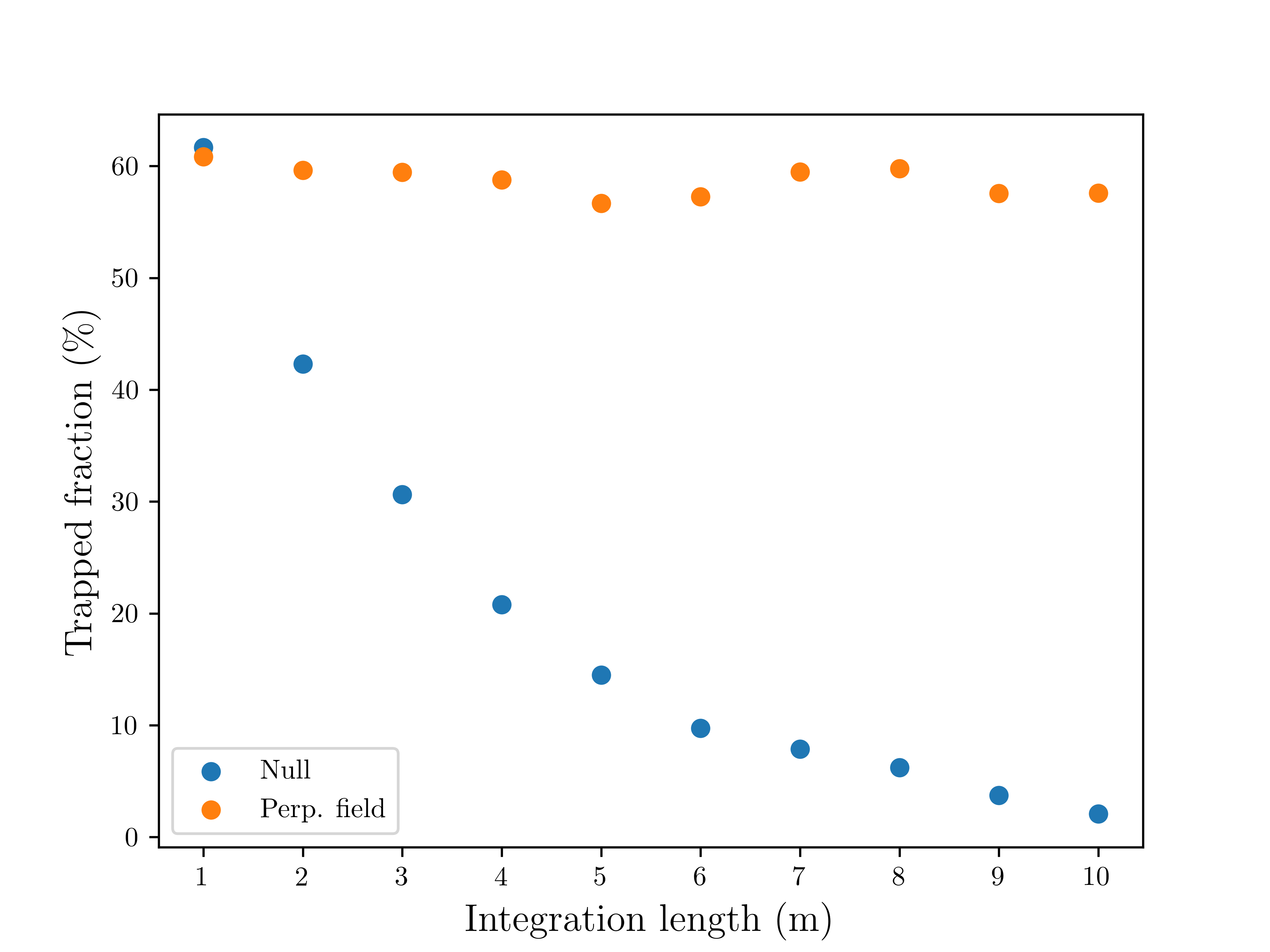}
    \caption{Trapping fraction (fraction of particles that have not reached a device wall) as a function of trajectory integration length. The perpendicular field has a trapping fraction of $\approx$60\%, independent of trajectory length, reflecting the stable trapped trajectories in a mirror configuration. The trapping fraction in the null configuration decreases with integration time, reflecting the unstable nature of trajectories close to the null. \label{fig:convergence}}
\end{figure}

As a means to quantify the trapping behavior of these two configurations, the simulation was repeated with varied integration length in order to observe the effect on apparent trapped fraction. The results are shown in figure \ref{fig:convergence}. The trapped fraction of the perpendicular field remains stable, indicating that these ions are the population trapped within the magnetic mirror. The apparently trapped fraction in the null configuration, however, decreases monotonically with integration length. This is due to the erratic trajectories of the ions,\hl{ }allowing them to eventually escape. 

\begin{figure}
    \centering
    \includegraphics[width=\linewidth]{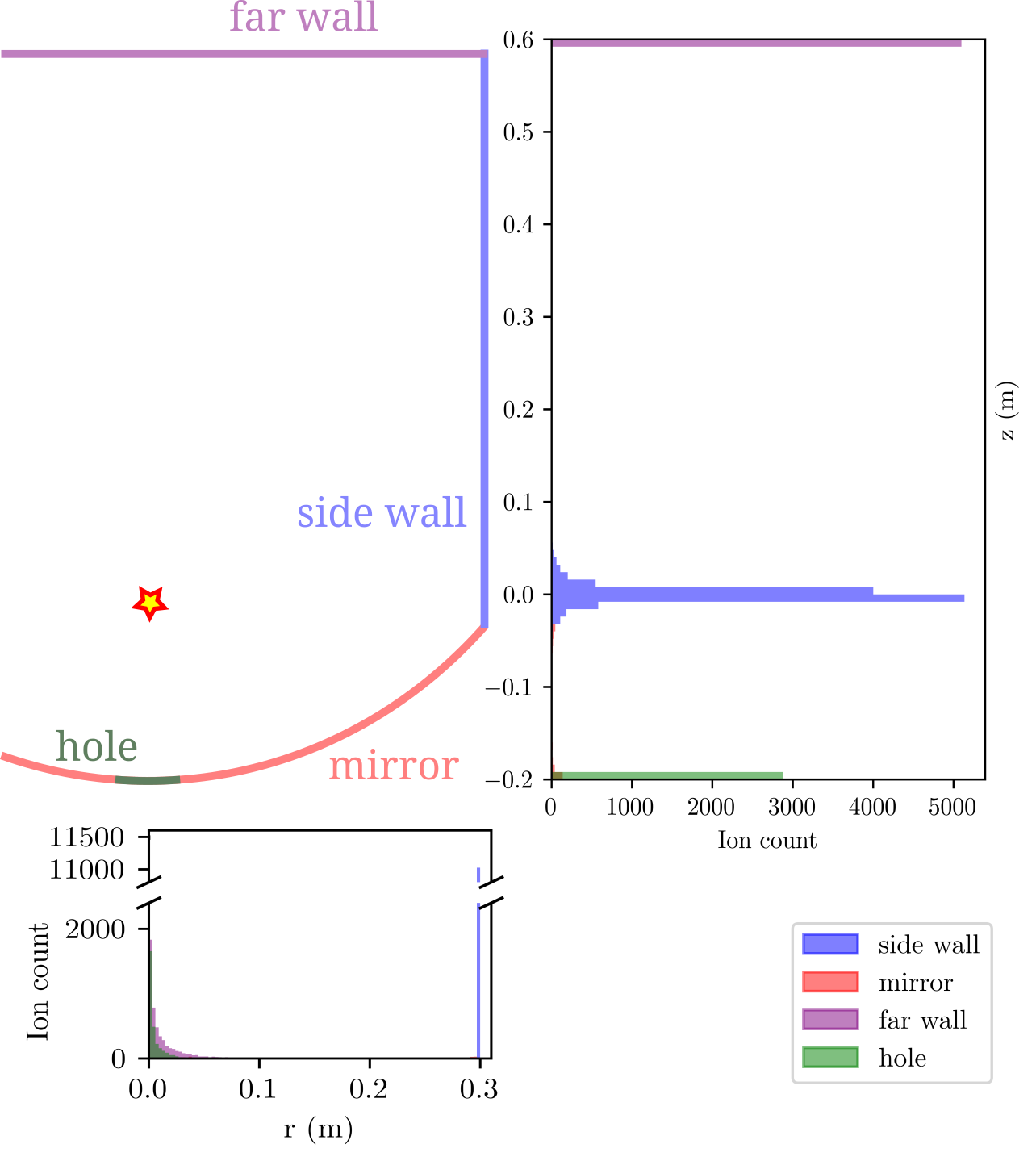}
    \caption{Distribution of ion strike points on the EUV system with the null geometry calculated with a $10$m integration length. The strike points are concentrated in the center of the top plate (purple), on the hole in the mirror (green), and in a narrow band on the side wall (blue). \label{fig:hist_null}}
    
\end{figure}

\begin{figure}
    \centering
    \includegraphics[width=\linewidth]{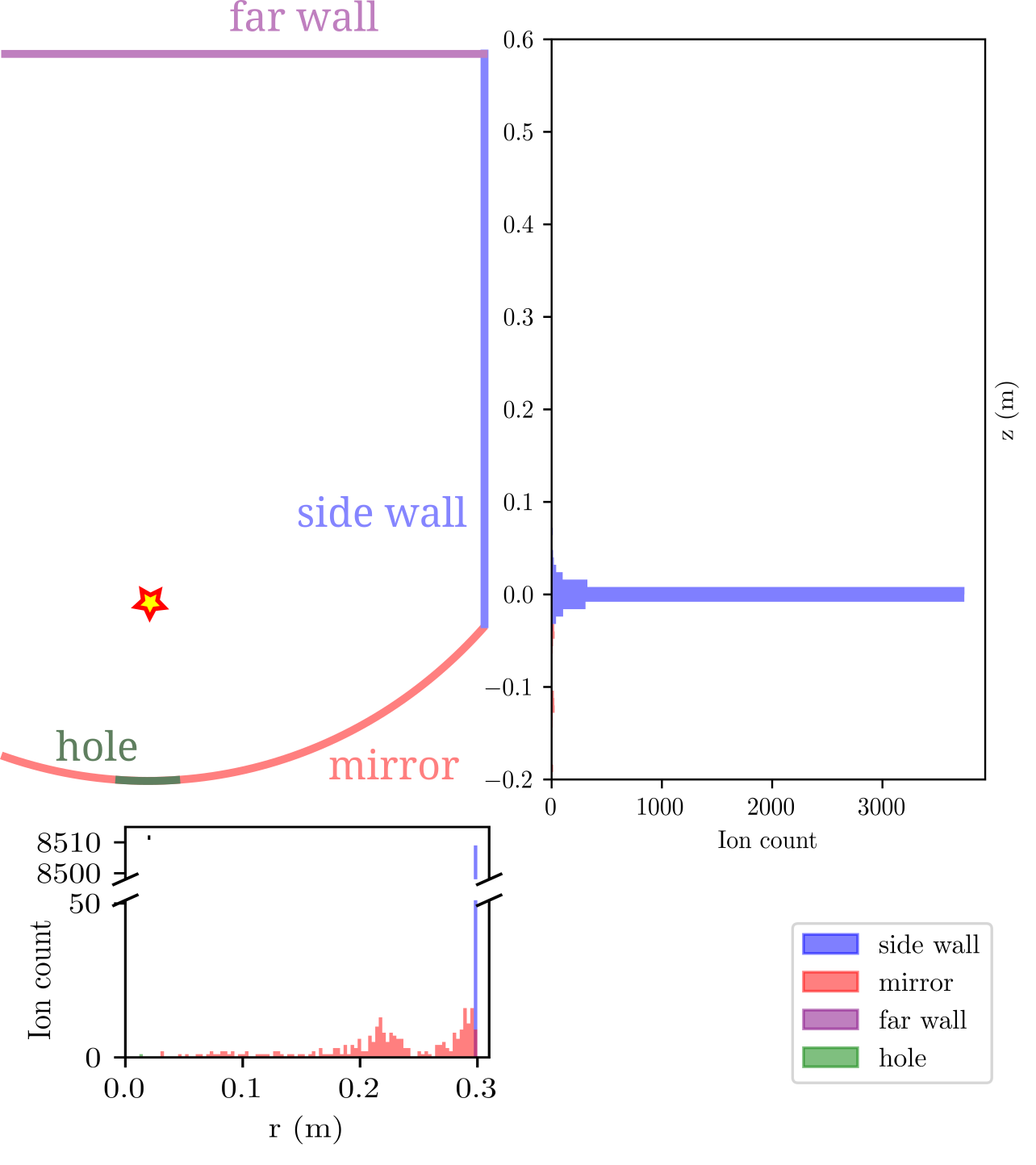}
    \caption{Distribution of ion strike points on the EUV system walls with the perpendicular field calculated with a $10$m integration length. The strike points are concentrated on on two spots just above the mirror. Furthermore, a significant fraction of the ions is trapped. These ions will not hit the mirror in this simulation\label{fig:hist_perp}}
    
\end{figure}

The distribution of ion deposition on the mirror and vessel wall for the compared configurations are shown in figures \ref{fig:hist_null} and \ref{fig:hist_perp}, and the resulting performance is summarized in table \ref{tab:results}. The null configuration directs ions along its fan plane, forming a ring around the side wall of the vessel, and along its spine, into the hole in the mirror or toward the center of the far wall of the vessel. The perpendicular field directs the ions towards two regions on opposite sides of the vessel side wall.



In this work we demonstrated that a magnetic null geometry is able to prevent the vast majority of tin ions from hitting the mirror in a realistically-dimensioned EUV system, without any trapping.
This geometry has previously been considered for confinement of fusion plasmas, but has been mostly abandoned due to inferior confinement - a benefit for debris mitigation. 
The null geometry was compared to the existing approach consisting of a perpendicular field. 
We showed that our\hl{ }approach could yield\hl{ }similar mitigation effectiveness results  i.e. 1 ion in 10$^5$ hitting the collector mirror. 
However, we find that magnetic nulls offer the unique additional capability of controlling the fraction of ions trapped within the system. 
Furthermore, a non-trapping configuration can be obtained. 

This non-trapping configuration has several potential benefits over the perpendicular field configuration: 
It can allow for operation with a much reduced background gas pressure and different composition, it reduces the ionization of the background gas induced by the fast ionic debris,  and it spreads the ion strikepoint distribution over a ring in the device. 

The perpendicular field requires a background gas to stop the trapped ions when they eventually neutralize, as they will then be unconfined.
With the null configuration, the calculus changes completely because all ions are on unstable orbits that eventually leave the system safely.  
A minimal gas pressure with a long mean-free-path between ion exchange events becomes feasible, and could help alleviate some of the issues caused by buffer gas: blistering due to ion implantation and significant EUV absorption by the gas. 
The optimum operational regime will likely still include minimal buffer gas that includes hydrogen, to clean the mirror via the formation of gaseous stannane, but this optimization is the subject of future study. 

The trapping fraction of the perpendicular field can be reduced by increasing the diameter of the coils, thus altering the mirror ratio $r_{\rm mirror}={B_{\rm max}}/{B_{\rm min}}$. For example, coils of 50 cm radius at the same locations produce a trapped fraction of 29\% (found numerically)\hl{.} 

In this work we only considered collisionless ion dynamics, and ignored the effect of electric fields, viscous drag, and charge exchange. 
This is because these processes depend heavily on background gas density and composition. 
Now that the null geometry has been shown to effectively protect the mirror from ion impingement, future work must focus on including these physical processes, and on finding an optimal mitigation strategy that includes buffer gas. 
Ion exchange and viscous drag can be incorporated using data from the ADAS library~\cite{summers2004adas}, and stochastic integration can be performed to include \hl{scattering off of buffer gas molecules.}, to determine the optimal buffer gas composition in the best-performing null configuration. 

This work was supported by the U.S. Department of Energy under contract number DE-AC02-09CH11466 (PPPL). \hl{The United States Government retains a non-exclusive, paid-up, irrevocable, world-wide license to publish or reproduce the published form of this manuscript, or allow others to do so, for United States Government purposes.} The authors were supported by the Laboratory Directed Research and Development (LDRD) Program of Princeton Plasma Physics Laboratory.

\bibliography{references}
\end{document}